\documentclass[twocolumn,pre,showpacs,amsmath,amssymb,reprint,superscriptaddress, nofootinbib]
{revtex4-1}

\usepackage{float}
\usepackage{graphicx}
\usepackage{amsmath,amssymb}
\usepackage{physics}
\usepackage[colorlinks=true, allcolors=blue]{hyperref}
\usepackage{amsmath}
\usepackage{mathtools}

\usepackage{array}
\usepackage{color}
\usepackage{multirow}
\usepackage{scalerel}
\usepackage{bm}

\setcitestyle{round}
\usepackage[dvipsnames]{xcolor}
\usepackage{comment}

\usepackage{lineno}

\usepackage{hyperref}
\hypersetup{
    colorlinks=true,
    linkcolor=blue,
    filecolor=blue,      
    urlcolor=blue,
    citecolor = blue,
    }

\usepackage{xcolor}

\makeatletter
\newcommand{\globalcolor}[1]{%
  \color{#1}\global\let\default@color\current@color
}
\makeatother

\begin{document}

\preprint{}

\title{Hamiltonian Dynamics and Structural States of Two-Dimensional Microswimmers}


\author{Yuval Shoham}
\affiliation{School of Physics and Astronomy and the Center for Physics and Chemistry of Living Systems, Tel Aviv University, Tel Aviv 6997801, Israel}
\author{Naomi Oppenheimer}
\email{naomiop@gmail.com}
\affiliation{School of Physics  and Astronomy and the Center for Physics and Chemistry of Living Systems, Tel Aviv University, Tel Aviv 6997801, Israel}

\date{\today}

\begin{abstract}

We show that a two-dimensional system of flocking microswimmers interacting hydrodynamically can be expressed using a Hamiltonian formalism. The Hamiltonian depends strictly on the angles between the particles and their swimming orientation, thereby restricting their available phase-space. Simulations of co-oriented microswimmers evolve into ``escalators” --- sharp lines at a particular tilt along which particles circulate. The conservation of the Hamiltonian and its symmetry germinate the self-assembly of the observed steady-state arrangements as confirmed by stability analysis.

\end{abstract}


\maketitle

At equilibrium, material structural states can be predicted and designed using an energetic description. Since structure encodes function, the energetic framework is a powerful tool in natural sciences. Yet, structural states are not prerogative of equilibrium --- structure also emerges in many-body systems, far from equilibrium, where canonical conservation laws fail \cite{vicsek1995novel}. For example, in Turing patterns \cite{turing1990chemical},  and in phase separation in biological and synthetic microswimmers \cite{palacci2013, cates2015motility, ben2022cooperation}.  In such systems, prediction becomes impossible without monitoring the full dynamical evolution of the many degrees of freedom, e.g., by agent-based simulations \cite{saintillan2007orientational,Nguyen2010, Lushi2014,saintillan2013active, manikantan2020tunable, lauga2021zigzag}, or a continuum description \cite{Toner2005, marchetti2013hydrodynamics, miles2019active}.
At equilibrium, finding the energy of a given state amounts to formulating its Hamiltonian, which is readily derived when the microscopic interactions are known. By contrast, even when the microscopic hydrodynamic interactions between active particles are known at great precision, an equivalent, general, Hamiltonian framework remains elusive.

\begin{figure}[tbh]
\centering
\includegraphics[width=0.45\textwidth]{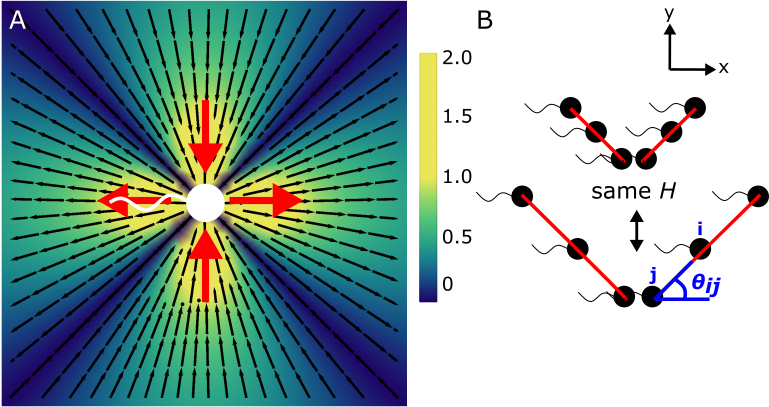} 
\caption{(a) Streamlines of a single particle swimming along the $x$ direction. Note the stagnation lines at $\pm \pi/4$, the inward flow along the vertical and outward along the horizontal. (b) Schematics showing six oriented swimmers, swimming in the $\hat{x}$ direction with the same Hamiltonian whether at close proximity or far apart as long as the angle between each two swimmers and their swimming direction, $\theta_{ij}$ is fixed.}
\label{fig_schematics}
\end{figure}

Here we show that for active particles in a 2D fluid, the equations of motion give rise to a geometric Hamiltonian description. We further show that when particles' orientations are aligned, such as in a flock, symmetries of the Hamiltonian limit the angular spread of the particles, resulting in an emergent structure of sharp lines at a given angle. This description applies to motile swimmers, such as bacteria \cite{Drescher2010, thery2020self}, and also to fixed active particles, such as proteins applying forces on the membrane \cite{oppenheimer2011plane}. An analogous geometric Hamiltonian proved useful for vortices in an ideal fluid \cite{Onsager1949, Newton2001, Aref1982}, and more recently also for active rotors in a viscous flow \cite{Lenz2003, lenz2004membranes, yeo2015collective, Lushi2015, Oppenheimer2019, oppenheimer2022hyperuniformity}, for sedimenting disk arrays  \cite{Chajwa2020, Chajwa2019, bolitho2022overdamped}, and for a swimmer or two interacting with a flow or an external field \cite{Zottl2012,  Stark2016, Bolitho2020}.
We consider hydrodynamic interactions between microscopic organisms such that inertia is negligible, and the governing equations are Stoke's equations. An active particle is force-free. Thus, to a leading order, it will generate a flow of a force-dipole given by 
$0=-\nabla p+\mu \nabla^2\vb{v}+\vb{D}\boldsymbol{:}\nabla \delta(\vb{r})$, where $\mu$ is the viscosity, $\vb{D}$ is the magnitude of the force dipole, and $\boldsymbol{:}$ is a double dot product. 
For an incompressible fluid ($\boldsymbol{\nabla}\cdot\vb{v}=0$) 
the resulting flow can be decomposed into an anti-symmetric part, a ``rotlet", and a symmetric part, a ``stresslet". They are compactly written in polar coordinates as
\begin{equation}
    \vb{v}=\frac{1}{2\pi r}\left[T \hat{\theta} + S\cos{(2\theta-2\phi)}\hat{r}\right],
    \label{eq:stresslet}
\end{equation}
where $T$ is the rotlet strength, $S$ the stresslet strength, $\phi$ the orientation of the force-dipole with respect to the $\hat{x}$ axis, and ${\bf r} = (x,y) = r(cos\theta, sin \theta)$ is the position of the particle. A general active particle will have a stresslet part, whether it is moving or statically applying active forces. Previous work focused on the rotational part \cite{Lenz2003, Lushi2014, Lushi2015, Oppenheimer2019, oppenheimer2022hyperuniformity}. Here we focus on the stresslet part of the flow-field. 
The incompressibility equation implies the existence of a vector potential such that $\vb{v}=\boldsymbol{\nabla}^{\perp}\psi$, where $\psi$ is the streamfunction, and   
$\boldsymbol{\nabla}^\perp\equiv\left(\partial_y, -\partial_x\right)$. These equations of motion are Hamilton equations, with $x$ and $y$ being the conjugate variables. 
The streamfunction of the symmetric part  of Eq.~\ref{eq:stresslet} is
$\psi_s = S\sin(2\theta - 2\phi)/\pi$ (see Fig.~\ref{fig_schematics}).

\begin{figure}[tbh]
\centering
\includegraphics[width=0.45\textwidth]{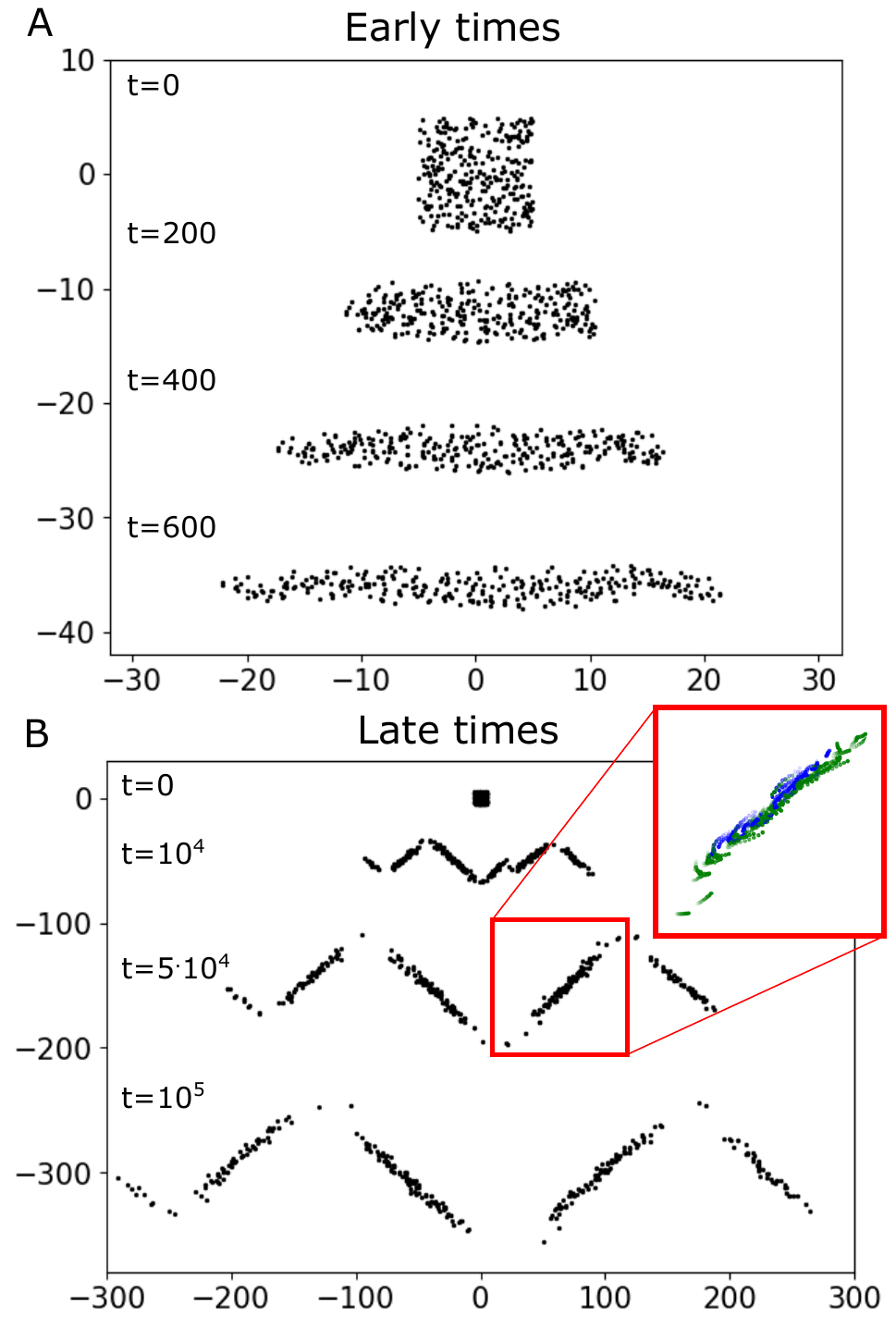} 
\caption{Snapshots from a molecular dynamics simulations of 300 swimmers, in a frame of reference moving with the particles. Particles are initiated randomly in a square of size $10\times 10$. Swimming directions are fixed along the $x$ axis. (a) Snapshots from the early times where the ensemble spreads and elongates. $y$ positions are shifted down between times for clarity. (b) Snapshots from later times where an instability is formed and grows. The particles develop sharp lines at $\pm \pi/4$, which we call escalators as particles circulate around them. Inset shows a zoom in on one of escalators with overlayed snapshots at different times. particles going left (right) are marked in blue (green).}
\label{fig_randomSnapshots}
\end{figure}

In a system of many similar active particles all swimming along the same direction with the same velocity, here chosen to be $\hat{x}$ (i.e. $\phi = 0$), where each particle is affected only by the other particle's flow-field, this streamfunction can be summed to become the Hamiltonian of the system.
The flow-field of the $i^{th}$ particle is given by,
\begin{eqnarray}
    \vb{v}_i = \sum_{j\neq i} \frac{S_j}{2\pi}\frac{\left(x_i-x_j\right)^2-\left(y_i-y_j\right)^2}{r_{ij}^4}\vb{r}_{ij},
    \label{eqVmany}
\end{eqnarray}
where $\vb{r}_{ij}\equiv \vb{r}_i-\vb{r}_j$ is the vector pointing from particle $j$ to particle $i$, and $S_i$ is the strength of the stresslet of the $i^{th}$ particle. These velocities can be derived from a Hamiltonian, $S_i \vb{v}_i = \nabla_i H$,
where $H$ is
\begin{eqnarray}
    H = \sum_{\substack{i,j , i\neq j}}\frac{S_i S_j}{2\pi}\sin{2\theta_{ij}}\label{Hamiltonian angle form},
\end{eqnarray}
and $\theta_{ij}$ is the relative angle between the vector connecting stresslets $i$ and $j$ and the $x$-axis (see Fig.~\ref{fig_schematics}B). We note four points about this Hamiltonian: (a) It does not depend on time, therefore, from Noether's theorem \cite{Noether1918}, it is conserved. 
(b) The Hamiltonian is scale-invariant, that is, it is the same whether the stresslets are very close or very far, as long as the relative angles between them are the same (see Fig.~\ref{fig_schematics}B). (c) It is symmetric with respect to $\pm \pi/4$. Therefore, we expect the solution to be symmetric around that angle.  (d) It is symmetric to translations, therefore ${\bf d}_{\rm act} \equiv \sum_i S_i {{\bf r}_i}/N = {\rm const}$, where ${\bf d}_{\rm act}$ is the ``center of activity" in  analogy to the center of mass. In what follows, we consider only stresslets with the same activity strength $S_i = S$ and find that a system of many oriented swimmers evolves into lines at angles $\pm \pi/4$ (see Fig.~\ref{fig_randomSnapshots}). To understand why that is, we start by examining the dynamics of two stresslets. 

A single stresslet does not move by the flow it creates. The simplest dynamic system is therefore composed of two particles, in which case, the Hamiltonian is reduced to $H = S^2\sin{2\theta}/(2\pi)$, and the relative angle itself is conserved. 
The dynamics of two stresslets are confined to the line connecting them.
When placed at initial distance $d_0$ and angle $\varphi$ relative to the $x$-axis, the relative distance between the two particles is $R = \sqrt{2S\cos{\left(2\varphi\right)}t/\pi + d_0^2}$. 
The angle $\varphi$ determines if the two stresslets collide or disperse. For $\varphi < \pi/4$, the stresslets repel each other. The repulsion scales with time as $R^2\sim t$, similar to diffusion. On the other hand, for $\varphi > \pi/4$, the stresslets collide after a finite time $t =d_0^2\pi/(2S\left|\cos{\left(2\varphi\right)}\right|)$. At exactly $\varphi~=~ \pi/4$, particles remain static, and the initial distance is fixed. 

\begin{figure}[tbh]
\centering
\includegraphics[width=0.45\textwidth]{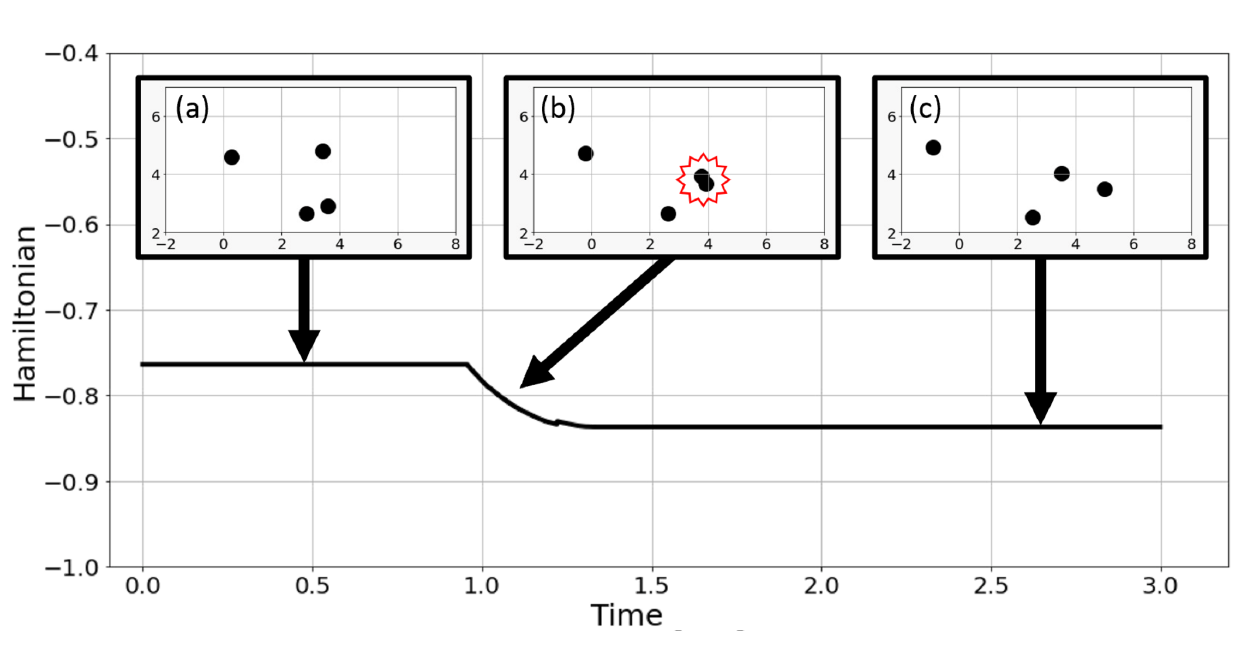} 
\caption{The Hamiltonian as a function of time for 4 random particles. The force dipole of all particles is oriented along the $\hat{x}$ axis.The Hamiltonian is conserved between collisions. A collision event is marked in red in the middle panel.}
\label{fig_Hamiltonian}
\end{figure}

A system of many stresslets can no longer be solved analytically. However, the conservation laws still apply. We numerically integrated Eq.~\ref{eqVmany}
using the python library \verb^scipy.integrate.DOP853^, which is an $8^{th}$ order Runge-Kutta method with an adaptive time stepper. 
The interaction between each two particles is radial and their interaction decays as $\sim r^{-1}$ (Eq.~\ref{eqVmany}). They repel or attract depending on their relative angle.  Thus, when two swimmers  attract, they accelerate toward each other and eventually collide, such that the velocity diverges. Actual active particles have a given size and cannot overlap. We, therefore, introduce soft steric repulsion of the form
$\Delta \vb{v}_{s} = 
\Delta t \, k_{s} \left(l_{s}-\left|\vb{r}\right|\right)\hat{r}$ if $ \left|\vb{r}\right|< l_{s}$ and zero otherwise. When two stresslets get closer than a certain steric length $l_{s}$, a repelling force proportional to their relative distance $\left|\vb{r}\right|$ by a very large spring constant $k_{s}$ is applied and pushes them apart. In our simulations, $l_{s}=0.001$ and $k_{s}=1,000$. This interaction is added to the regular stresslet interaction. Due to the steric interactions, the Hamiltonian is no longer strictly conserved. To ensure that collisions do not dominate the behavior of the system, we work in the dilute limit where collisions are rare, with average distances between particles much larger than the steric length. We verified that the Hamiltonian is accurately conserved in between collisions (see Fig.~\ref{fig_Hamiltonian}).

    

We initialize 300 oriented stresslets randomly positioned in a square and let them evolve, all swimming in the $\hat{x}$ direction. At first, the system follows the general shape of a single stresslet shown in Fig.~\ref{fig_schematics}A--- the system compresses in the $y$-direction and expands in the $x$-direction, into a ``street" of stresslets, see Fig.~\ref{fig_randomSnapshots}A. At intermediate times, this street shows instabilities, which grow over time. At long times, these instabilities create stable shapes where particles concentrate along inclined streets at angles $\pm \varphi=\pi/4$ (see Fig.~\ref{fig_randomSnapshots}B), which we term ``escalators". There is circulation around these escalators, with particles above and below going in opposite directions. See inset in Fig.~\ref{fig_randomSnapshots}B, for overlayed snapshots at preceding times where particles are color coded according to their direction of motion. 

Why are there always escalators of $\varphi=\pm \pi/4$ when the system is initialized in a random square? We explain this using symmetry and Hamiltonian conservation arguments.
We begin by showing that the system can only evolve into several escalators and not a single one.
The Hamiltonian of the initial state is, on average, zero since the initial angles are random and $H$ can be written as $H = N (N-1) S^2 \left<\sin{2\theta_{ij}}\right>/2\pi = 0$, where $\left<.\right>$ is the average over the ensemble, $N$ being the number of stresslets. 
The Hamiltonian of an escalator with $\varphi = \pm \pi/4$, on the other hand, is non-zero. In fact it has maximal magnitude.  
Given that the Hamiltonian is conserved, a system of stresslets scattered in a square cannot develop into a single escalator. Indeed, the system always decomposes into a few escalators, each with an inclination angle $\varphi = \pm \pi/4$.
A simple case for this decomposition is two escalators with opposing inclination angles $\varphi=\pm \pi/4$, placed at opposing sides of the $y$ axis in symmetric form, see Fig.~\ref{fig_schematics}B. Mirroring the system along the $y$-axis, the Hamiltonian is $H^{\text{mirror}_y} \propto \left<\sin{\left(2\cdot\left(\pi-\theta_{ij}\right)\right)}\right> = -H$,
which implies again that the Hamiltonian is zero. 
Thus, a symmetric combination of stresslets conserves the Hamiltonian. 
Moreover, there is no limitation to the number of escalators the system can develop into, and different runs resulted in different numbers.
The Hamiltonian is symmetric around $\pm \pi/4$, so the steady state configuration needs to exhibit this symmetry. We go on to test the stability of particles aligned at different angles. 

\textit{Stability Analysis For a Street of stresslets}.
 \begin{figure}
     \centering
     \includegraphics[width=0.5\textwidth]{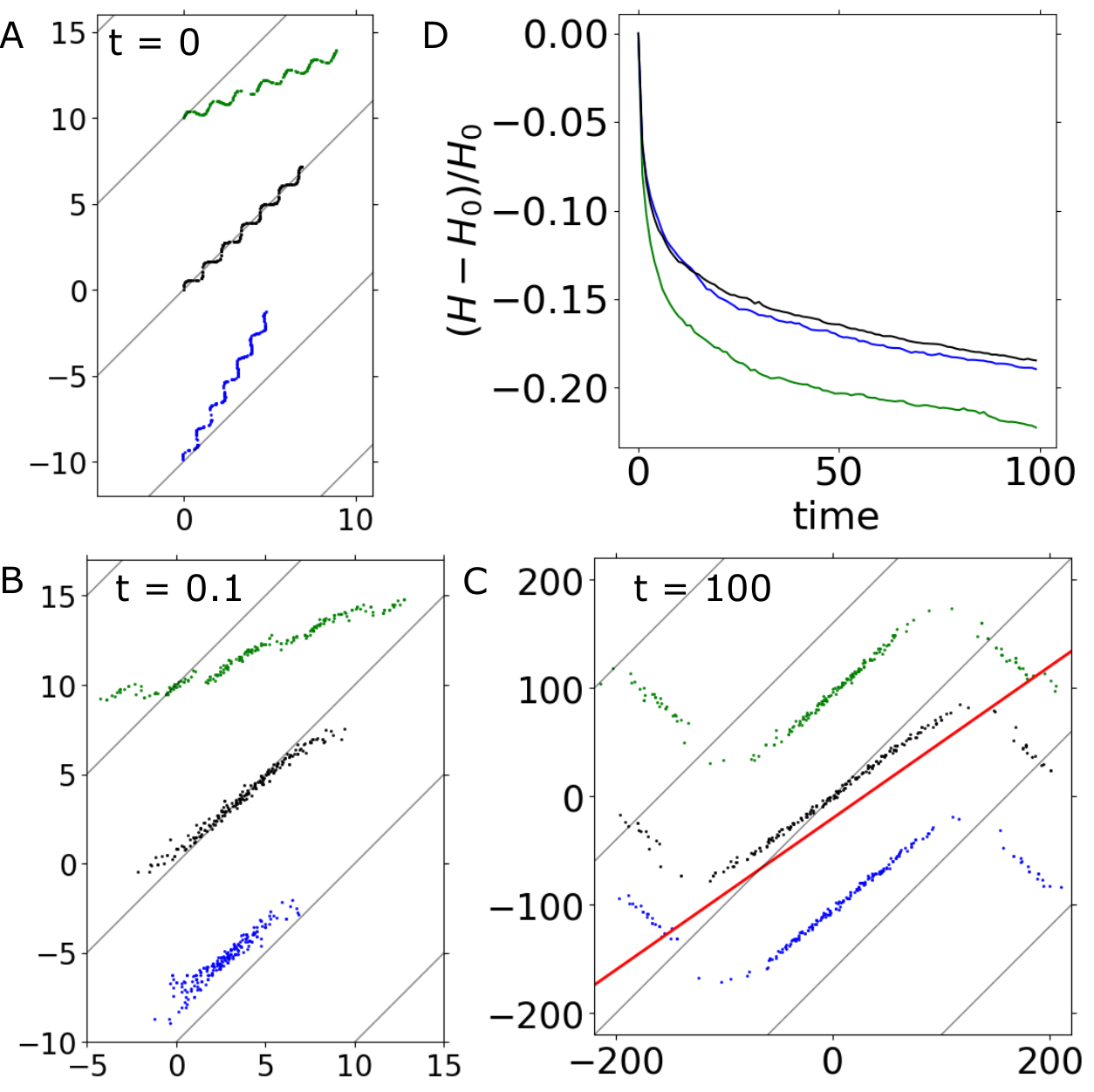}
     \caption{Stability of particles at different angles under perturbation. We initiate 200 particles randomly on a line tilted at three different angles: $\pi/8$ in green, $\pi/4$ in black and $\pi/3$ in green. We add a small sine-wave perturbation to their positions and let the system evolve over time. (a) Initial positions, (b) Configuration after a short period ($t = 0.1$) shows all initial conditions resulted in escelators at about $\pi/4$ degrees. Note how the $\pi/8$ street breaks into smaller streets. (c) Result at long times with all initial conditions resulting in lines at $\sim 40^{\circ}$. Slightly less than $\pi/4$ and consistent with around $15\%$ decrease in the value of the Hamitlonian. At later times the system continues to spread but maintains these angles as there are hardly any collisions. (d) The relative error of the Hamitlonian as a function of time, where $H_0$ is its initial value. }
     \label{fig_escalator_stability}
 \end{figure}
We use three different methods to test the stability of an escalator, inclined at an angle $\varphi$. First, we use linear stability analysis and show that the stability of the inclined escalator is of non-linear nature, as the first order of perturbation gives a fixed point of type center. 
Next, we use numeric simulations of escalators with different inclination angles to show that escalators are unstable --- except when $\varphi= \pi/4$. Lastly, by calculating analytically the velocity field created by a continuous distribution of particles in an escalator, we show that for $\varphi< \pi/4$, there is a repelling force pushing particles away. In addition, escalators at $\varphi> \pi/4$ have an attractive force that causes a collapse and are therefore, inherently unstable. We combine the results to conclude that the only stable escalator is $\varphi= \pm \pi/4$.

We follow the method of linear stability analysis used in Ref.~\cite{Saffman1993} \S8. We start with a continuous line of stresslets with strength density $S$. In complex notation the velocity is
\begin{eqnarray}
    \dot{Z} = \frac{S}{2\pi}\frac{1}{2Z}\left(1+\left(\frac{Z}{\bar{Z}}\right)^2\right),
\end{eqnarray}
where $Z=x+iy$, and $\bar{Z}=x-iy$. Along the parametric line $\left\{Z_s=Z\left(s\right)|s\in\left(-\infty,\infty\right)\right\}$, the velocity at each point $\dot{Z}_s$ is
\begin{eqnarray}
    \dot{Z}_s = \int_{-\infty}^\infty\frac{S}{4\pi \left(Z_s-Z_{s^\prime}\right)}\left(1+\left(\frac{Z_s-Z_{s^\prime}}{\bar{Z}_s-\bar{Z}_{s^\prime}}\right)^2\right)\mathrm{d}s^\prime\label{velocity integral}.
\end{eqnarray}
For a stresslet ``street" with an inclination $\varphi$ compared to the $x$-axis, $Z\left(s\right)=se^{i\varphi}$. Adding a small Fourier-decomposed perturbation $\varepsilon_s= \sum_q a_qe^{iqs}$ to the street, keeping only its lowest non-zero order, the equations of motion for $Z$ and $\bar{Z}$ (Eq.~\ref{velocity integral}) become

\begin{equation}
    \frac{\mathrm{d}}{\mathrm{d}t}
    \begin{pmatrix}
    a_q\\
    \Bar{a}_{-q}
    \end{pmatrix} = -\frac{Sq}{2}
    \begin{pmatrix}
    i\sin{4\varphi} & -e^{4i\varphi}\\
    e^{-4i\varphi} & i\sin{4\varphi}
    \end{pmatrix}
    \begin{pmatrix}
    a_q\\
    \Bar{a}_{-q}
    \end{pmatrix},
\end{equation}
with eigenvalues 
$\lambda_{1,2} = -\frac{1}{2}iSq\left(\sin{4\varphi}\pm1\right)$.
Note that the eigenvalues are imaginary and create a neutrally stable point of type center. 
Therefore, in the linear limit, small perturbations will not grow nor decay. A similar calculation in a 3D fluid showed that stresslets in that case are linearly unstable \cite{lauga2021zigzag}. We next show by two other means that streets at all angles except $\pm  \pi/4$ are, in fact, unstable due to higher-order perturbations.
When a street of any inclination is infinite, and the stresslets are at equal distances apart, the system is completely stationary. On the other hand, when a street is finite, only an escalator with inclination $\pm \pi/4$ will remain stationary, because of the stagnation lines at $\pm  \pi/4$. This already hints that an unperturbed escalator of inclination $\pm \pi/4$ is more stable than any other unperturbed street.

Next, we use numeric simulations of perturbed escalators. We initialize 200 stresslets in random locations along tilted lines at a fixed angle and perturb them by adding a sine wave to their locations in the perpendicular direction. We then let the system develop over time.
After a short time ($t=0.1$), escalators with an inclination angle $\varphi\neq \pi/4$ are broken and end up as a system of many escalators approaching $\varphi = \pi/4$ (Fig.~\ref{fig_escalator_stability}). 
The result at long times ($t\sim 100$, Fig.~\ref{fig_escalator_stability}C) is, though, not exactly $ \pi/4$, due to a decrease in the Hamiltonian (see Fig.~\ref{fig_escalator_stability}D). Such a decrease in the Hamiltonian comes from the steric interactions introduced to the system (Fig.~\ref{fig_schematics}C), which are necessary to avoid the divergence of the velocity field at short distances since $v \sim 1/r$. From its initial value, the Hamiltonian loses $\sim 15\%$ at $t=100$. This change seems to scale sub-linearly with time, i.e. $\leq \sqrt{t}$. In a way, the system diffuses into another state due to collisions. 
Let us attempt to take into account the decrease in the Hamiltonian and estimate the angle of a ``perfect" escalator with the same decreased Hamiltonian. The resulting angle is aroung $40^{\circ}$ and agrees well with the simulation predictions, as shown in the red dashed line in Fig.~\ref{fig_escalator_stability}.

To test analytically the stability, we use a continuous model of an infinite escalator where particles are equispaced and ask what flow field it creates --- if its flow field repels particles away from it, it is certainly not stable.
We sum an infinite series of stresslets laid on a line of inclination $\varphi$ distanced $L$ apart. We calculate the velocity that the system of stresslets creates at the time of initiation.
Instead of inclining the street, we look at stresslets on the $x$-axis whose direction is rotated by $\varphi$. The velocity due to a single stresslet is given by the radial part of Eq.~\ref{eq:stresslet}.
The velocity at a distance $h$ above the escalator is given by an infinite sum over all stresslets which gives
${\bf v} =  \left[ \left(\coth{\rho} - \rho\csch^2{\rho}\right)\sin{2\varphi}, \rho\csch^2{\rho}\cos{2\varphi} \right]S/2L $
with $\rho = \pi h/L$.
The $y$-component represents the instability of the line. 
We see there will be a repelling force from the inclined street when $\cos{2\varphi}>0$. This means that for $\cos{2\varphi}>0$ (i.e. $\varphi < \pi/4$),  the inclined street is unstable to perturbations. For $\cos{2\varphi} = 0$, that is $\varphi = \pi/4$, the velocity outside the escalator is zero, therefore a perturbation neither grows nor decays. What happens for angles larger than $\pi/4$? In that case, as we have seen for two stressletss, there is an attraction between the particles. Because there are no repelling forces other than collisions, a stresslet street at such an angle collapses onto itself and loses its stability. We finally conclude that indeed $\varphi=\pi/4$ is the only stable angle.

\textit{Discussion}.
In this work, we introduced a new method to describe a 2D system of microswimmers robustly, using multipole expansion and a Hamiltonian formalism, which can be used to unveil the dynamics of colonies of microbiological swimmers and other synthetic active systems. The symmetries of the Hamiltonian limit the possible steady state configurations of the particles. Such a Hamiltonian description has been useful in the study of vortices in an ideal fluid, and here we find its
application in a viscosity dominated, active, many-body system. 
Numeric simulations gave further insight into the dynamics. 
We found that a system of co-aligned, yet randomly positioned swimmers progressively flattened into a linear street, before the onset of an instability where the swimmers self-assembled into ``escalators" in which particles circulate on canted conveyor belts.
In future directions of this work, we plan to look into populations of active particles with different orientations, thus altering the Hamiltonian, and possibly changing the dynamics drastically.  

\textit{Acknowledgments}.
We thank Haim Diamant for fruitful discussions. This research was supported by the ISRAEL SCIENCE FOUNDATION (grant No. 1752/20). 


\bibliography{library}

\begin{thebibliography}{34}%
\makeatletter
\providecommand \@ifxundefined [1]{%
 \@ifx{#1\undefined}
}%
\providecommand \@ifnum [1]{%
 \ifnum #1\expandafter \@firstoftwo
 \else \expandafter \@secondoftwo
 \fi
}%
\providecommand \@ifx [1]{%
 \ifx #1\expandafter \@firstoftwo
 \else \expandafter \@secondoftwo
 \fi
}%
\providecommand \natexlab [1]{#1}%
\providecommand \enquote  [1]{``#1''}%
\providecommand \bibnamefont  [1]{#1}%
\providecommand \bibfnamefont [1]{#1}%
\providecommand \citenamefont [1]{#1}%
\providecommand \href@noop [0]{\@secondoftwo}%
\providecommand \href [0]{\begingroup \@sanitize@url \@href}%
\providecommand \@href[1]{\@@startlink{#1}\@@href}%
\providecommand \@@href[1]{\endgroup#1\@@endlink}%
\providecommand \@sanitize@url [0]{\catcode `\\12\catcode `\$12\catcode
  `\&12\catcode `\#12\catcode `\^12\catcode `\_12\catcode `\%12\relax}%
\providecommand \@@startlink[1]{}%
\providecommand \@@endlink[0]{}%
\providecommand \url  [0]{\begingroup\@sanitize@url \@url }%
\providecommand \@url [1]{\endgroup\@href {#1}{\urlprefix }}%
\providecommand \urlprefix  [0]{URL }%
\providecommand \Eprint [0]{\href }%
\providecommand \doibase [0]{http://dx.doi.org/}%
\providecommand \selectlanguage [0]{\@gobble}%
\providecommand \bibinfo  [0]{\@secondoftwo}%
\providecommand \bibfield  [0]{\@secondoftwo}%
\providecommand \translation [1]{[#1]}%
\providecommand \BibitemOpen [0]{}%
\providecommand \bibitemStop [0]{}%
\providecommand \bibitemNoStop [0]{.\EOS\space}%
\providecommand \EOS [0]{\spacefactor3000\relax}%
\providecommand \BibitemShut  [1]{\csname bibitem#1\endcsname}%
\let\auto@bib@innerbib\@empty
\bibitem [{\citenamefont {Vicsek}\ \emph {et~al.}(1995)\citenamefont {Vicsek},
  \citenamefont {Czir{\'o}k}, \citenamefont {Ben-Jacob}, \citenamefont
  {Cohen},\ and\ \citenamefont {Shochet}}]{vicsek1995novel}%
  \BibitemOpen
  \bibfield  {author} {\bibinfo {author} {\bibfnamefont {T.}~\bibnamefont
  {Vicsek}}, \bibinfo {author} {\bibfnamefont {A.}~\bibnamefont {Czir{\'o}k}},
  \bibinfo {author} {\bibfnamefont {E.}~\bibnamefont {Ben-Jacob}}, \bibinfo
  {author} {\bibfnamefont {I.}~\bibnamefont {Cohen}}, \ and\ \bibinfo {author}
  {\bibfnamefont {O.}~\bibnamefont {Shochet}},\ }\href@noop {} {\bibfield
  {journal} {\bibinfo  {journal} {Physical review letters}\ }\textbf {\bibinfo
  {volume} {75}},\ \bibinfo {pages} {1226} (\bibinfo {year}
  {1995})}\BibitemShut {NoStop}%
\bibitem [{\citenamefont {Turing}(1990)}]{turing1990chemical}%
  \BibitemOpen
  \bibfield  {author} {\bibinfo {author} {\bibfnamefont {A.~M.}\ \bibnamefont
  {Turing}},\ }\href@noop {} {\bibfield  {journal} {\bibinfo  {journal}
  {Bulletin of mathematical biology}\ }\textbf {\bibinfo {volume} {52}},\
  \bibinfo {pages} {153} (\bibinfo {year} {1990})}\BibitemShut {NoStop}%
\bibitem [{\citenamefont {Palacci}\ \emph {et~al.}(3222)\citenamefont
  {Palacci}, \citenamefont {Sacanna}, \citenamefont {Steinberg}, \citenamefont
  {Pine},\ and\ \citenamefont {Chaikin}}]{palacci2013}%
  \BibitemOpen
  \bibfield  {author} {\bibinfo {author} {\bibfnamefont {J.}~\bibnamefont
  {Palacci}}, \bibinfo {author} {\bibfnamefont {S.}~\bibnamefont {Sacanna}},
  \bibinfo {author} {\bibfnamefont {A.~P.}\ \bibnamefont {Steinberg}}, \bibinfo
  {author} {\bibfnamefont {D.~J.}\ \bibnamefont {Pine}}, \ and\ \bibinfo
  {author} {\bibfnamefont {P.~M.}\ \bibnamefont {Chaikin}},\ }\href {\doibase
  10.1126/science.1230020} {\bibfield  {journal} {\bibinfo  {journal}
  {Science.}\ }\textbf {\bibinfo {volume} {339}},\ \bibinfo {pages} {936}
  (\bibinfo {year} {2013222})}\BibitemShut {NoStop}%
\bibitem [{\citenamefont {Cates}\ and\ \citenamefont
  {Tailleur}(2015)}]{cates2015motility}%
  \BibitemOpen
  \bibfield  {author} {\bibinfo {author} {\bibfnamefont {M.~E.}\ \bibnamefont
  {Cates}}\ and\ \bibinfo {author} {\bibfnamefont {J.}~\bibnamefont
  {Tailleur}},\ }\href@noop {} {\bibfield  {journal} {\bibinfo  {journal}
  {Annu. Rev. Condens. Matter Phys.}\ }\textbf {\bibinfo {volume} {6}},\
  \bibinfo {pages} {219} (\bibinfo {year} {2015})}\BibitemShut {NoStop}%
\bibitem [{\citenamefont {Ben~Zion}\ \emph {et~al.}(2022)\citenamefont
  {Ben~Zion}, \citenamefont {Caba}, \citenamefont {Modin},\ and\ \citenamefont
  {Chaikin}}]{ben2022cooperation}%
  \BibitemOpen
  \bibfield  {author} {\bibinfo {author} {\bibfnamefont {M.~Y.}\ \bibnamefont
  {Ben~Zion}}, \bibinfo {author} {\bibfnamefont {Y.}~\bibnamefont {Caba}},
  \bibinfo {author} {\bibfnamefont {A.}~\bibnamefont {Modin}}, \ and\ \bibinfo
  {author} {\bibfnamefont {P.~M.}\ \bibnamefont {Chaikin}},\ }\href@noop {}
  {\bibfield  {journal} {\bibinfo  {journal} {Nature Communications}\ }\textbf
  {\bibinfo {volume} {13}},\ \bibinfo {pages} {184} (\bibinfo {year}
  {2022})}\BibitemShut {NoStop}%
\bibitem [{\citenamefont {Saintillan}\ and\ \citenamefont
  {Shelley}(2007)}]{saintillan2007orientational}%
  \BibitemOpen
  \bibfield  {author} {\bibinfo {author} {\bibfnamefont {D.}~\bibnamefont
  {Saintillan}}\ and\ \bibinfo {author} {\bibfnamefont {M.~J.}\ \bibnamefont
  {Shelley}},\ }\href@noop {} {\bibfield  {journal} {\bibinfo  {journal}
  {Physical review letters}\ }\textbf {\bibinfo {volume} {99}},\ \bibinfo
  {pages} {058102} (\bibinfo {year} {2007})}\BibitemShut {NoStop}%
\bibitem [{\citenamefont {Nguyen}\ \emph {et~al.}(2010)\citenamefont {Nguyen},
  \citenamefont {Atkinson}, \citenamefont {Park}, \citenamefont {Maclennan},
  \citenamefont {Glaser},\ and\ \citenamefont {Clark}}]{Nguyen2010}%
  \BibitemOpen
  \bibfield  {author} {\bibinfo {author} {\bibfnamefont {Z.~H.}\ \bibnamefont
  {Nguyen}}, \bibinfo {author} {\bibfnamefont {M.}~\bibnamefont {Atkinson}},
  \bibinfo {author} {\bibfnamefont {C.~S.}\ \bibnamefont {Park}}, \bibinfo
  {author} {\bibfnamefont {J.}~\bibnamefont {Maclennan}}, \bibinfo {author}
  {\bibfnamefont {M.}~\bibnamefont {Glaser}}, \ and\ \bibinfo {author}
  {\bibfnamefont {N.}~\bibnamefont {Clark}},\ }\href {\doibase
  10.1103/PhysRevLett.105.268304} {\bibfield  {journal} {\bibinfo  {journal}
  {Physical Review Letters}\ }\textbf {\bibinfo {volume} {105}},\ \bibinfo
  {pages} {268304} (\bibinfo {year} {2010})}\BibitemShut {NoStop}%
\bibitem [{\citenamefont {Lushi}\ \emph {et~al.}(2014)\citenamefont {Lushi},
  \citenamefont {Wioland},\ and\ \citenamefont {Goldstein}}]{Lushi2014}%
  \BibitemOpen
  \bibfield  {author} {\bibinfo {author} {\bibfnamefont {E.}~\bibnamefont
  {Lushi}}, \bibinfo {author} {\bibfnamefont {H.}~\bibnamefont {Wioland}}, \
  and\ \bibinfo {author} {\bibfnamefont {R.~E.}\ \bibnamefont {Goldstein}},\
  }\href {\doibase 10.1073/pnas.1405698111} {\bibfield  {journal} {\bibinfo
  {journal} {Proceedings of the National Academy of Sciences}\ }\textbf
  {\bibinfo {volume} {111}},\ \bibinfo {pages} {9733} (\bibinfo {year}
  {2014})}\BibitemShut {NoStop}%
\bibitem [{\citenamefont {Saintillan}\ and\ \citenamefont
  {Shelley}(2013)}]{saintillan2013active}%
  \BibitemOpen
  \bibfield  {author} {\bibinfo {author} {\bibfnamefont {D.}~\bibnamefont
  {Saintillan}}\ and\ \bibinfo {author} {\bibfnamefont {M.~J.}\ \bibnamefont
  {Shelley}},\ }\href@noop {} {\bibfield  {journal} {\bibinfo  {journal}
  {Comptes Rendus Physique}\ }\textbf {\bibinfo {volume} {14}},\ \bibinfo
  {pages} {497} (\bibinfo {year} {2013})}\BibitemShut {NoStop}%
\bibitem [{\citenamefont {Manikantan}(2020)}]{manikantan2020tunable}%
  \BibitemOpen
  \bibfield  {author} {\bibinfo {author} {\bibfnamefont {H.}~\bibnamefont
  {Manikantan}},\ }\href@noop {} {\bibfield  {journal} {\bibinfo  {journal}
  {Physical review letters}\ }\textbf {\bibinfo {volume} {125}},\ \bibinfo
  {pages} {268101} (\bibinfo {year} {2020})}\BibitemShut {NoStop}%
\bibitem [{\citenamefont {Lauga}\ \emph {et~al.}(2021)\citenamefont {Lauga},
  \citenamefont {Dang},\ and\ \citenamefont {Ishikawa}}]{lauga2021zigzag}%
  \BibitemOpen
  \bibfield  {author} {\bibinfo {author} {\bibfnamefont {E.}~\bibnamefont
  {Lauga}}, \bibinfo {author} {\bibfnamefont {T.~N.}\ \bibnamefont {Dang}}, \
  and\ \bibinfo {author} {\bibfnamefont {T.}~\bibnamefont {Ishikawa}},\
  }\href@noop {} {\bibfield  {journal} {\bibinfo  {journal} {Europhysics
  Letters}\ }\textbf {\bibinfo {volume} {133}},\ \bibinfo {pages} {44002}
  (\bibinfo {year} {2021})}\BibitemShut {NoStop}%
\bibitem [{\citenamefont {Toner}\ \emph {et~al.}(2005)\citenamefont {Toner},
  \citenamefont {Tu},\ and\ \citenamefont {Ramaswamy}}]{Toner2005}%
  \BibitemOpen
  \bibfield  {author} {\bibinfo {author} {\bibfnamefont {J.}~\bibnamefont
  {Toner}}, \bibinfo {author} {\bibfnamefont {Y.}~\bibnamefont {Tu}}, \ and\
  \bibinfo {author} {\bibfnamefont {S.}~\bibnamefont {Ramaswamy}},\ }\href
  {\doibase 10.1016/j.aop.2005.04.011} {\bibfield  {journal} {\bibinfo
  {journal} {Annals of Physics}\ }\textbf {\bibinfo {volume} {318}},\ \bibinfo
  {pages} {170} (\bibinfo {year} {2005})}\BibitemShut {NoStop}%
\bibitem [{\citenamefont {Marchetti}\ \emph {et~al.}(2013)\citenamefont
  {Marchetti}, \citenamefont {Joanny}, \citenamefont {Ramaswamy}, \citenamefont
  {Liverpool}, \citenamefont {Prost}, \citenamefont {Rao},\ and\ \citenamefont
  {Simha}}]{marchetti2013hydrodynamics}%
  \BibitemOpen
  \bibfield  {author} {\bibinfo {author} {\bibfnamefont {M.~C.}\ \bibnamefont
  {Marchetti}}, \bibinfo {author} {\bibfnamefont {J.-F.}\ \bibnamefont
  {Joanny}}, \bibinfo {author} {\bibfnamefont {S.}~\bibnamefont {Ramaswamy}},
  \bibinfo {author} {\bibfnamefont {T.~B.}\ \bibnamefont {Liverpool}}, \bibinfo
  {author} {\bibfnamefont {J.}~\bibnamefont {Prost}}, \bibinfo {author}
  {\bibfnamefont {M.}~\bibnamefont {Rao}}, \ and\ \bibinfo {author}
  {\bibfnamefont {R.~A.}\ \bibnamefont {Simha}},\ }\href@noop {} {\bibfield
  {journal} {\bibinfo  {journal} {Reviews of modern physics}\ }\textbf
  {\bibinfo {volume} {85}},\ \bibinfo {pages} {1143} (\bibinfo {year}
  {2013})}\BibitemShut {NoStop}%
\bibitem [{\citenamefont {Miles}\ \emph {et~al.}(2019)\citenamefont {Miles},
  \citenamefont {Evans}, \citenamefont {Shelley},\ and\ \citenamefont
  {Spagnolie}}]{miles2019active}%
  \BibitemOpen
  \bibfield  {author} {\bibinfo {author} {\bibfnamefont {C.~J.}\ \bibnamefont
  {Miles}}, \bibinfo {author} {\bibfnamefont {A.~A.}\ \bibnamefont {Evans}},
  \bibinfo {author} {\bibfnamefont {M.~J.}\ \bibnamefont {Shelley}}, \ and\
  \bibinfo {author} {\bibfnamefont {S.~E.}\ \bibnamefont {Spagnolie}},\
  }\href@noop {} {\bibfield  {journal} {\bibinfo  {journal} {Physical Review
  Letters}\ }\textbf {\bibinfo {volume} {122}},\ \bibinfo {pages} {098002}
  (\bibinfo {year} {2019})}\BibitemShut {NoStop}%
\bibitem [{\citenamefont {Drescher}\ \emph {et~al.}(2010)\citenamefont
  {Drescher}, \citenamefont {Goldstein}, \citenamefont {Michel}, \citenamefont
  {Polin},\ and\ \citenamefont {Tuval}}]{Drescher2010}%
  \BibitemOpen
  \bibfield  {author} {\bibinfo {author} {\bibfnamefont {K.}~\bibnamefont
  {Drescher}}, \bibinfo {author} {\bibfnamefont {R.~E.}\ \bibnamefont
  {Goldstein}}, \bibinfo {author} {\bibfnamefont {N.}~\bibnamefont {Michel}},
  \bibinfo {author} {\bibfnamefont {M.}~\bibnamefont {Polin}}, \ and\ \bibinfo
  {author} {\bibfnamefont {I.}~\bibnamefont {Tuval}},\ }\href {\doibase
  10.1103/PhysRevLett.105.168101} {\bibfield  {journal} {\bibinfo  {journal}
  {Physical Review Letters}\ }\textbf {\bibinfo {volume} {105}},\ \bibinfo
  {pages} {168101} (\bibinfo {year} {2010})}\BibitemShut {NoStop}%
\bibitem [{\citenamefont {Th{\'e}ry}\ \emph {et~al.}(2020)\citenamefont
  {Th{\'e}ry}, \citenamefont {Le~Nagard}, \citenamefont {Ono-dit Biot},
  \citenamefont {Fradin}, \citenamefont {Dalnoki-Veress},\ and\ \citenamefont
  {Lauga}}]{thery2020self}%
  \BibitemOpen
  \bibfield  {author} {\bibinfo {author} {\bibfnamefont {A.}~\bibnamefont
  {Th{\'e}ry}}, \bibinfo {author} {\bibfnamefont {L.}~\bibnamefont
  {Le~Nagard}}, \bibinfo {author} {\bibfnamefont {J.-C.}\ \bibnamefont {Ono-dit
  Biot}}, \bibinfo {author} {\bibfnamefont {C.}~\bibnamefont {Fradin}},
  \bibinfo {author} {\bibfnamefont {K.}~\bibnamefont {Dalnoki-Veress}}, \ and\
  \bibinfo {author} {\bibfnamefont {E.}~\bibnamefont {Lauga}},\ }\href@noop {}
  {\bibfield  {journal} {\bibinfo  {journal} {Scientific reports}\ }\textbf
  {\bibinfo {volume} {10}},\ \bibinfo {pages} {1} (\bibinfo {year}
  {2020})}\BibitemShut {NoStop}%
\bibitem [{\citenamefont {Oppenheimer}\ and\ \citenamefont
  {Diamant}(2011)}]{oppenheimer2011plane}%
  \BibitemOpen
  \bibfield  {author} {\bibinfo {author} {\bibfnamefont {N.}~\bibnamefont
  {Oppenheimer}}\ and\ \bibinfo {author} {\bibfnamefont {H.}~\bibnamefont
  {Diamant}},\ }\href@noop {} {\bibfield  {journal} {\bibinfo  {journal}
  {Physical review letters}\ }\textbf {\bibinfo {volume} {107}},\ \bibinfo
  {pages} {258102} (\bibinfo {year} {2011})}\BibitemShut {NoStop}%
\bibitem [{\citenamefont {Onsager}(1949)}]{Onsager1949}%
  \BibitemOpen
  \bibfield  {author} {\bibinfo {author} {\bibfnamefont {L.}~\bibnamefont
  {Onsager}},\ }\href {\doibase 10.1007/BF02780991} {\bibfield  {journal}
  {\bibinfo  {journal} {Il Nuovo Cimento}\ }\textbf {\bibinfo {volume} {6}},\
  \bibinfo {pages} {279} (\bibinfo {year} {1949})}\BibitemShut {NoStop}%
\bibitem [{\citenamefont {Newton}(2001)}]{Newton2001}%
  \BibitemOpen
  \bibfield  {author} {\bibinfo {author} {\bibfnamefont {P.~K.}\ \bibnamefont
  {Newton}},\ }\href {\doibase 10.1007/978-1-4684-9290-3} {\emph {\bibinfo
  {title} {The N-Vortex Problem}}},\ Vol.\ \bibinfo {volume} {145}\ (\bibinfo
  {publisher} {Springer New York},\ \bibinfo {year} {2001})\BibitemShut
  {NoStop}%
\bibitem [{\citenamefont {Aref}\ and\ \citenamefont
  {Pomphrey}(1982)}]{Aref1982}%
  \BibitemOpen
  \bibfield  {author} {\bibinfo {author} {\bibfnamefont {H.}~\bibnamefont
  {Aref}}\ and\ \bibinfo {author} {\bibfnamefont {N.}~\bibnamefont
  {Pomphrey}},\ }\href {\doibase 10.1098/rspa.1982.0047} {\bibfield  {journal}
  {\bibinfo  {journal} {Proceedings of the Royal Society of London. A.
  Mathematical and Physical Sciences}\ }\textbf {\bibinfo {volume} {380}},\
  \bibinfo {pages} {359} (\bibinfo {year} {1982})}\BibitemShut {NoStop}%
\bibitem [{\citenamefont {Lenz}\ \emph {et~al.}(2003)\citenamefont {Lenz},
  \citenamefont {Joanny}, \citenamefont {Jülicher},\ and\ \citenamefont
  {Prost}}]{Lenz2003}%
  \BibitemOpen
  \bibfield  {author} {\bibinfo {author} {\bibfnamefont {P.}~\bibnamefont
  {Lenz}}, \bibinfo {author} {\bibfnamefont {J.-F.}\ \bibnamefont {Joanny}},
  \bibinfo {author} {\bibfnamefont {F.}~\bibnamefont {Jülicher}}, \ and\
  \bibinfo {author} {\bibfnamefont {J.}~\bibnamefont {Prost}},\ }\href
  {\doibase 10.1103/PhysRevLett.91.108104} {\bibfield  {journal} {\bibinfo
  {journal} {Physical Review Letters}\ }\textbf {\bibinfo {volume} {91}},\
  \bibinfo {pages} {108104} (\bibinfo {year} {2003})}\BibitemShut {NoStop}%
\bibitem [{\citenamefont {Lenz}\ \emph {et~al.}(2004)\citenamefont {Lenz},
  \citenamefont {Joanny}, \citenamefont {J{\"u}licher},\ and\ \citenamefont
  {Prost}}]{lenz2004membranes}%
  \BibitemOpen
  \bibfield  {author} {\bibinfo {author} {\bibfnamefont {P.}~\bibnamefont
  {Lenz}}, \bibinfo {author} {\bibfnamefont {J.-F.}\ \bibnamefont {Joanny}},
  \bibinfo {author} {\bibfnamefont {F.}~\bibnamefont {J{\"u}licher}}, \ and\
  \bibinfo {author} {\bibfnamefont {J.}~\bibnamefont {Prost}},\ }\href@noop {}
  {\bibfield  {journal} {\bibinfo  {journal} {The European Physical Journal E}\
  }\textbf {\bibinfo {volume} {13}},\ \bibinfo {pages} {379} (\bibinfo {year}
  {2004})}\BibitemShut {NoStop}%
\bibitem [{\citenamefont {Yeo}\ \emph {et~al.}(2015)\citenamefont {Yeo},
  \citenamefont {Lushi},\ and\ \citenamefont {Vlahovska}}]{yeo2015collective}%
  \BibitemOpen
  \bibfield  {author} {\bibinfo {author} {\bibfnamefont {K.}~\bibnamefont
  {Yeo}}, \bibinfo {author} {\bibfnamefont {E.}~\bibnamefont {Lushi}}, \ and\
  \bibinfo {author} {\bibfnamefont {P.~M.}\ \bibnamefont {Vlahovska}},\
  }\href@noop {} {\bibfield  {journal} {\bibinfo  {journal} {Physical review
  letters}\ }\textbf {\bibinfo {volume} {114}},\ \bibinfo {pages} {188301}
  (\bibinfo {year} {2015})}\BibitemShut {NoStop}%
\bibitem [{\citenamefont {Lushi}\ and\ \citenamefont
  {Vlahovska}(2015)}]{Lushi2015}%
  \BibitemOpen
  \bibfield  {author} {\bibinfo {author} {\bibfnamefont {E.}~\bibnamefont
  {Lushi}}\ and\ \bibinfo {author} {\bibfnamefont {P.~M.}\ \bibnamefont
  {Vlahovska}},\ }\href {\doibase 10.1007/s00332-015-9254-9} {\bibfield
  {journal} {\bibinfo  {journal} {Journal of Nonlinear Science}\ }\textbf
  {\bibinfo {volume} {25}},\ \bibinfo {pages} {1111} (\bibinfo {year}
  {2015})}\BibitemShut {NoStop}%
\bibitem [{\citenamefont {Oppenheimer}\ \emph {et~al.}(2019)\citenamefont
  {Oppenheimer}, \citenamefont {Stein},\ and\ \citenamefont
  {Shelley}}]{Oppenheimer2019}%
  \BibitemOpen
  \bibfield  {author} {\bibinfo {author} {\bibfnamefont {N.}~\bibnamefont
  {Oppenheimer}}, \bibinfo {author} {\bibfnamefont {D.~B.}\ \bibnamefont
  {Stein}}, \ and\ \bibinfo {author} {\bibfnamefont {M.~J.}\ \bibnamefont
  {Shelley}},\ }\href {\doibase 10.1103/PhysRevLett.123.148101} {\bibfield
  {journal} {\bibinfo  {journal} {Physical Review Letters}\ }\textbf {\bibinfo
  {volume} {123}},\ \bibinfo {pages} {148101} (\bibinfo {year}
  {2019})}\BibitemShut {NoStop}%
\bibitem [{\citenamefont {Oppenheimer}\ \emph {et~al.}(2022)\citenamefont
  {Oppenheimer}, \citenamefont {Stein}, \citenamefont {Zion},\ and\
  \citenamefont {Shelley}}]{oppenheimer2022hyperuniformity}%
  \BibitemOpen
  \bibfield  {author} {\bibinfo {author} {\bibfnamefont {N.}~\bibnamefont
  {Oppenheimer}}, \bibinfo {author} {\bibfnamefont {D.~B.}\ \bibnamefont
  {Stein}}, \bibinfo {author} {\bibfnamefont {M.~Y.~B.}\ \bibnamefont {Zion}},
  \ and\ \bibinfo {author} {\bibfnamefont {M.~J.}\ \bibnamefont {Shelley}},\
  }\href@noop {} {\bibfield  {journal} {\bibinfo  {journal} {Nature
  communications}\ }\textbf {\bibinfo {volume} {13}},\ \bibinfo {pages} {804}
  (\bibinfo {year} {2022})}\BibitemShut {NoStop}%
\bibitem [{\citenamefont {Chajwa}\ \emph {et~al.}(2020)\citenamefont {Chajwa},
  \citenamefont {Menon}, \citenamefont {Ramaswamy},\ and\ \citenamefont
  {Govindarajan}}]{Chajwa2020}%
  \BibitemOpen
  \bibfield  {author} {\bibinfo {author} {\bibfnamefont {R.}~\bibnamefont
  {Chajwa}}, \bibinfo {author} {\bibfnamefont {N.}~\bibnamefont {Menon}},
  \bibinfo {author} {\bibfnamefont {S.}~\bibnamefont {Ramaswamy}}, \ and\
  \bibinfo {author} {\bibfnamefont {R.}~\bibnamefont {Govindarajan}},\ }\href
  {\doibase 10.1103/PhysRevX.10.041016} {\bibfield  {journal} {\bibinfo
  {journal} {Physical Review X}\ }\textbf {\bibinfo {volume} {10}},\ \bibinfo
  {pages} {041016} (\bibinfo {year} {2020})}\BibitemShut {NoStop}%
\bibitem [{\citenamefont {Chajwa}\ \emph {et~al.}(2019)\citenamefont {Chajwa},
  \citenamefont {Menon},\ and\ \citenamefont {Ramaswamy}}]{Chajwa2019}%
  \BibitemOpen
  \bibfield  {author} {\bibinfo {author} {\bibfnamefont {R.}~\bibnamefont
  {Chajwa}}, \bibinfo {author} {\bibfnamefont {N.}~\bibnamefont {Menon}}, \
  and\ \bibinfo {author} {\bibfnamefont {S.}~\bibnamefont {Ramaswamy}},\ }\href
  {\doibase 10.1103/PhysRevLett.122.224501} {\bibfield  {journal} {\bibinfo
  {journal} {Physical Review Letters}\ }\textbf {\bibinfo {volume} {122}},\
  \bibinfo {pages} {224501} (\bibinfo {year} {2019})}\BibitemShut {NoStop}%
\bibitem [{\citenamefont {Bolitho}\ and\ \citenamefont
  {Adhikari}(2022)}]{bolitho2022overdamped}%
  \BibitemOpen
  \bibfield  {author} {\bibinfo {author} {\bibfnamefont {A.}~\bibnamefont
  {Bolitho}}\ and\ \bibinfo {author} {\bibfnamefont {R.}~\bibnamefont
  {Adhikari}},\ }\href@noop {} {\bibfield  {journal} {\bibinfo  {journal}
  {arXiv preprint arXiv:2202.06102}\ } (\bibinfo {year} {2022})}\BibitemShut
  {NoStop}%
\bibitem [{\citenamefont {Zöttl}\ and\ \citenamefont
  {Stark}(2012)}]{Zottl2012}%
  \BibitemOpen
  \bibfield  {author} {\bibinfo {author} {\bibfnamefont {A.}~\bibnamefont
  {Zöttl}}\ and\ \bibinfo {author} {\bibfnamefont {H.}~\bibnamefont {Stark}},\
  }\href {\doibase 10.1103/PhysRevLett.108.218104} {\bibfield  {journal}
  {\bibinfo  {journal} {Physical Review Letters}\ }\textbf {\bibinfo {volume}
  {108}},\ \bibinfo {pages} {218104} (\bibinfo {year} {2012})}\BibitemShut
  {NoStop}%
\bibitem [{\citenamefont {Stark}(2016)}]{Stark2016}%
  \BibitemOpen
  \bibfield  {author} {\bibinfo {author} {\bibfnamefont {H.}~\bibnamefont
  {Stark}},\ }\href {\doibase 10.1140/epjst/e2016-60060-2} {\bibfield
  {journal} {\bibinfo  {journal} {The European Physical Journal Special
  Topics}\ }\textbf {\bibinfo {volume} {225}},\ \bibinfo {pages} {2369}
  (\bibinfo {year} {2016})}\BibitemShut {NoStop}%
\bibitem [{\citenamefont {Bolitho}\ \emph {et~al.}(2020)\citenamefont
  {Bolitho}, \citenamefont {Singh},\ and\ \citenamefont
  {Adhikari}}]{Bolitho2020}%
  \BibitemOpen
  \bibfield  {author} {\bibinfo {author} {\bibfnamefont {A.}~\bibnamefont
  {Bolitho}}, \bibinfo {author} {\bibfnamefont {R.}~\bibnamefont {Singh}}, \
  and\ \bibinfo {author} {\bibfnamefont {R.}~\bibnamefont {Adhikari}},\ }\href
  {\doibase 10.1103/PhysRevLett.124.088003} {\bibfield  {journal} {\bibinfo
  {journal} {Physical Review Letters}\ }\textbf {\bibinfo {volume} {124}},\
  \bibinfo {pages} {088003} (\bibinfo {year} {2020})}\BibitemShut {NoStop}%
\bibitem [{\citenamefont {Noether}(1918)}]{Noether1918}%
  \BibitemOpen
  \bibfield  {author} {\bibinfo {author} {\bibfnamefont {E.}~\bibnamefont
  {Noether}},\ }\href {http://eudml.org/doc/59024} {\bibfield  {journal}
  {\bibinfo  {journal} {Nachrichten von der Gesellschaft der Wissenschaften zu
  Göttingen, Mathematisch-Physikalische Klasse}\ }\textbf {\bibinfo {volume}
  {1918}},\ \bibinfo {pages} {235} (\bibinfo {year} {1918})}\BibitemShut
  {NoStop}%
\bibitem [{\citenamefont {Saffman}(1993)}]{Saffman1993}%
  \BibitemOpen
  \bibfield  {author} {\bibinfo {author} {\bibfnamefont {P.~G.}\ \bibnamefont
  {Saffman}},\ }\href {\doibase 10.1017/CBO9780511624063} {\emph {\bibinfo
  {title} {Vortex Dynamics}}}\ (\bibinfo  {publisher} {Cambridge University
  Press},\ \bibinfo {year} {1993})\BibitemShut {NoStop}%
\end{thebibliography}%

\end{document}